# An improved smart meta-superconductor $MgB_2$

Xiaopeng Zhao*, Qingyu Hai, Miao Shi, Honggang Chen, Yongbo Li and Yao Qi

Smart Materials Laboratory, Department of Applied Physics, Northwestern Polytechnical University, Xi'an 710129, China. *Correspondence: xpzhao@nwpu.edu.cn (X.Z.)

**Abstract:** Increasing and improving the critical transition temperature (Tc), current density (Jc) and Meissner effect (Hc) of conventional superconductors are the most important problems in superconductivity research, but progress has been slow for many years. In this study, by introducing the p-n junction electroluminescent inhomogeneous phase with red wavelength to realize energy injection, we found the improved property of smart meta-superconductors $MgB_2$, the critical transition temperature Tc increases by 0.8K, the current density Jc increases by 37%, and the diamagnetism of Meissner effect Hc also significantly improved, compared with pure $MgB_2$. Compared with previous yttrium oxide inhomogeneous phase, p-n junction has higher luminescence intensity, longer stable life and simpler external field requirements. The coupling between superconducting electrons and surface plasmon polaritons may be explain this phenomenon. The realization of smart meta-superconductor by this electroluminescent inhomogeneous phase provides a new way to improve the performance of superconductors.

## 1. Introduction

Superconductivity has greatly promoted the progress of industrial technology since its discovery, and also expanded people's understanding of condensed matter physics [1]. The superconducting materials have a wide range of applications, such as electric grids that can transmit power without energy loss [2], and quantum computing devices that use superposition and quantum entanglement to perform calculations [3, 4]. The pursuit of superconducting materials with high critical temperature Tc has been promoting the research. High-temperature superconductor [5, 6], iron-based superconductor [7, 8], high-pressure superconductor [9-12] and photoinduced superconductor [13, 14] have been gradually studied and discovered. Superconductors have zero resistance characteristics and complete diamagnetism (Meissner effect) [15-17]. Therefore, the transition from superconducting state to non-superconducting state has characteristic parameters: critical transition temperature ($T_C$), critical current density ($J_C$) and critical magnetic field ($H_C$) [15, 18]. In recent years, it has been found that the hydride has a high transition temperature under high pressure. At 155 GPa [9], the superconductivity of the sulfur hydride system is 203 K, and at 267 GPa [12], the superconductivity of the carbonized sulfur hydride system at room temperature is 287.7 K. Although this method can achieve higher superconducting transition temperatures and even room temperature superconductivity, the extremely high pressure and small sample size limit its further applications.

The discovery of $MgB_2$ superconductor [19] soon aroused great interest in the scientific community due to its excellent superconductivity and simple preparation process, especially high Tc. In order to improve the superconductivity of $MgB_2$, various methods have been adopted [20-24], which can not only improve the practical application of $MgB_2$, but also further clarify its superconductivity mechanism. Chemical doping is often used to study superconductivity. Unfortunately, many experimental results have confirmed that this approach reduces Tc of $MgB_2$ [25-30]. So far, there is no effective strategy to improve Tc of $MgB_2$. Chemical doping is the simplest method to change $J_C$ of superconductor. Doping graphene in $MgB_2$ [31], and $Al_2O_3$ [32] and $MgO$ [33] in BiSrCaCuO will reduce $J_C$ under zero magnetic field. The same time, adding anthracene into $MgB_2$ [34] and $Cr_2O_3$ into BiSrCaCuO [35] will increase $J_C$ under zero magnetic field. Under zero magnetic field, chemical doping increases or decreases $J_C$ of superconductor, but correspondingly decreases $T_C$. There is no particularly effective method to increase $T_C$ and $J_C$ at the same time.

Metamaterials are composite materials with artificial structures. It has supernormal physical properties that natural materials do not have, and these supernormal properties are determined by special artificial structures [36-38]. In 2007, we proposed to introduce inorganic ZnO electroluminescent (EL) material into Bi(Pb)SrCaCuO superconductor at high temperature to affect the superconducting transition temperature of Bi(Pb)SrCaCuO [39,40]. In recent years, $MgB_2$ and Bi(Pb)SrCaCuO smart meta-superconductors have been constructed, SMSCs consist of superconducting particles and $Y_2O_3:Eu^{3+}$ or $Y_2O_3:Eu^{3+}+Ag$ luminescent inhomogeneous phases. We doped $Y_2O_3:Eu^{3+}$ and $Y_2O_3:Eu^{3+}+Ag$ materials in conventional $MgB_2$ and high temperature Bi(Pb)SrCaCuO superconductors[40-47]. In this method, the electroluminescent (EL) material is directly doped into the superconductor to form a smart meta-superconductor. When the Tc of SMSCs is measured by the four-probe method, the external electric field can stimulate the inhomogeneous phase to produce EL, which can achieve the purpose of strengthening the Cooper pair and lead to the macroscopic change of Tc. SMSCs is a material that can adjust and improve Tc through external field stimulation, which is a new property that cannot be achieved by traditional second phase doping [44-46]. The results show that doping of $Y_2O_3:Eu^{3+}$ and $Y_2O_3:Eu^{3+}+Ag$ EL materials increases $T_C$ of $MgB_2$ and Bi(Pb)SrCaCuO. The maximum change of transition temperature of $MgB_2$ is 1.2 K[46], the maximum change of the zero resistance temperature of Bi(Pb)SrCaCuO increased by 4 K, and the initial transition temperature increased by 6.3 K [45]. We believe that this is because superconducting particles acting as microelectrodes excite the inhomogeneous phase EL under the action of an applied electric field, and the energy injection promotes the formation of electron pairs. Therefore, $T_C$ of $MgB_2$ and Bi(Pb)SrCaCuO can be increased by EL [43,44]. Recently, $J_C$ and Meissner effects of $MgB_2$ and Bi(Pb)SrCaCuO smart meta-superconductors have been investigated[47]. The results show that the addition of $Y_2O_3:Eu^{3+}$ and $Y_2O_3:Eu^{3+}+Ag$ increases $T_C$ and $J_C$ of $MgB_2$ and Bi(Pb)SrCaCuO. $J_C$ of luminescent inhomogeneous phase doped samples decreases to a minimum at higher temperatures. Dc magnetization data show that the doping of $Y_2O_3$:$Eu^{3+}$ and $Y_2O_3$:$Eu^{3+}$+Ag inhomogeneous phase leads to the Meissner effect of $MgB_2$ and Bi(Pb)SrCaCuO at higher temperatures, while non-luminescent doping reduces the Meissner effect temperature. Through previous studies, it has been confirmed that the rare earth oxide inhomogeneous phase can improve the superconducting critical transition temperature $T_C$, critical current density $J_C$ and magnetic field $H_C$ of conventional superconductor and high temperature oxide, and improve the complete diamagnetism (Meissner effect). However, it is very difficult to improve the electroluminescence intensity of rare earth oxides, short luminescence life, large applied electric field and other factors, which limit the improvement of superconducting electric performance.

In this paper, the smart superconductivity of $MgB_2$ was studied by introducing p-n junction electroluminescence inhomogeneous phase to realize energy injection and improve electron pairing. Studies show that the high luminescence intensity and long life of p-n junction can ensure the stability of material properties. In addition, p-n junction excitation is easier, only a few volts of excitation, not hundreds or even thousands of volts, the electric field applied by the four-point method for measuring superconductivity can be satisfied. Because p-n junction inhomogeneous phase exhibits good behavior under field excitation, the optimum amount of dopable inhomogeneous phase increased from 0.5% to 1% compared to oxide. Therefore, the critical temperature Tc, current density Jc and Meissner effect Hc of superconducting transition are higher than those of oxide inhomogeneous phase. In particular, the performance stability of the material has been greatly improved, and can be stable for more than several hundred hours. We hold the opinion that the photons generated by the inhomogeneous phase of p-n junction electroluminescence interact with some superconducting electrons to generate surface plasmon polaritons (SPPs) promote electron pair transport. The smart superconductivity of p-n electroluminescent inhomogeneous phase provides a new way to improve the performance of superconductor.

## 2. Model

Figure 1 shows the $MgB_2$ SMSCs model constructed with polycrystalline $MgB_2$ as raw material. Gray polyhedron are polycrystalline $MgB_2$ particles, Φ is the particle size of $MgB_2$ particle, which will be described in detail in the experimental part. The red particles are p-n junction particles with red light wavelengths, which are dispersed among $MgB_2$ particles as inhomogeneous phase. The introduction of inhomogeneous phase inevitably reduces the Tc of $MgB_2$, mainly because the doped inhomogeneous phase is not a superconductor, which is detrimental to the superconductivity of $MgB_2$, such as the MgO impurity phase in $MgB_2$. For convenience, the reduction of Tc after the introduction of dopants is called impurity effect [39-41]. The incorporation of inhomogeneous phases has been proved to be an effective method to improve Tc in $MgB_2$ and Bi(Pb)SrCaCuO systems. For example, the introduction of electroluminescence $Y_2O_3$:$Eu^{3+}$ and $Y_2O_3$:$Eu^{3+}$+Ag can produce electroluminescence effect and increase Tc [42-46]. There is obvious competition between impurity effect and EL excitation effect of inhomogeneous phase. When EL excitation effect is dominant, Tc is improved (ΔTc > 0). Otherwise, inhomogeneous phase is introduced to reduce Tc (ΔTc < 0). Therefore, the impurity effect should be reduced as much as possible and EL excitation effect should be enhanced to obtain high Tc samples. To obtain a smart meta-superconductor with superconductivity, its Tc can be improved and adjusted by adding EL inhomogeneous phase [45, 46]. It has been known that variations in Tc are often related to variations in electron density [48, 59]. However, under the current preparation conditions, the inhomogeneous phase only exists between $MgB_2$ particles and does not react with $MgB_2$. Moreover, the diffusion between the inhomogeneous phase and $MgB_2$ particles is difficult, and the electron density cannot be significantly changed. Therefore, electron density is not a key tuning parameter affecting Tc variation. During the measurement process, the applied electric field forms a local electric field in the superconductor and excites the inhomogeneous phase to generate EL excited photon injection energy, which is beneficial to the enhancement of Cooper pair and the change of Tc. However, given that photons may destroy the Cooper pair, the mechanism for the occurrence of Tc changes needs to be further explored. Later, according to the experimental results, we will explain this phenomenon by the inhomogeneous phase EL.

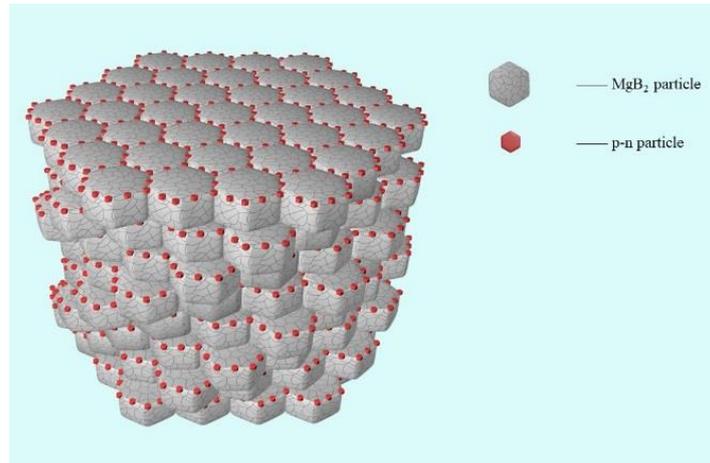

Figure 1. $MgB_2$ SMSCs model diagram

## 3. Experiment

### *3.1 Preparation of p-n junction luminescent particles*

Purchase commercial red LED epitaxial chip, its luminous composition is AlGaInP composite structure, luminous wavelength is 623nm. Compound AlGaInP is produced by Shandong Huaguang Optoelectronics Co., LTD. China. It uses trimethylgallium, trimethylindium, trimethylaluminum and phosphoane as raw materials and as reactants, which are brought into the vacuum furnace by hydrogen for reaction growth on the substrate. The melting point of the product AlGaInP is between 1200-1400℃. We strip compounds off the substrate, grind them to get particles, shape as shown in figure 3b, about 4x4x3 μm particles. The electroluminescence test method is the same as that of rare earth luminescent particles, and the measurement conditions were given in the text. Applied voltage <10V, current <10mA, test luminescence curve is shown in Figure 2. The luminescence curves of the electroluminescent rare earth oxide particles in the figure were obtained from the samples prepared by our group [50, 51]. It can be seen that the luminescence intensity of p-n junction particles is much higher than that of electroluminescent rare earth oxide particles, especially under more than 2000 hours of work, the luminescence intensity almost did not decay, and the luminescence behavior did not change after 100 days. The characteristics of high strength and long life of p-n junction provide a solid foundation for improving smart meta-superconductors.

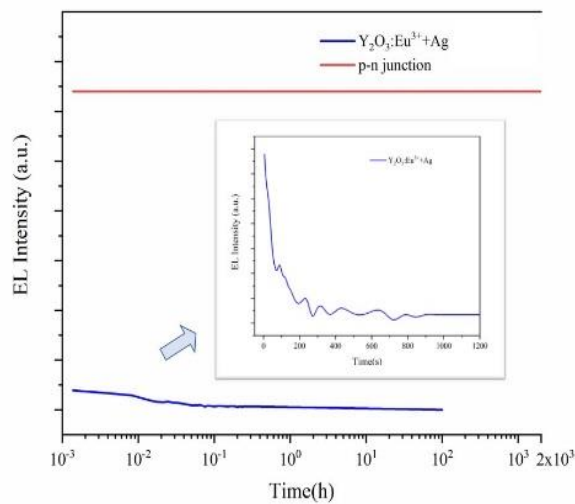

Figure 2. Luminescence intensity and lifetime test curves of p-n junction particles and rare earth oxide particles in red light wavelengths

### 3.2. Preparation of MgB$_2$ superconductor and inhomogeneous phase samples

Magnesium diboride (MgB$_2$) is purchased from Alfa Aesar with a purity of 99% and particle size of 100 mesh. A certain amount of MgB$_2$ basic powder raw material was put into a 500 mesh stainless steel standard sieve, and the large size particles were removed by screening to get the basic uniform size of MgB$_2$ particles, diameter $\Phi \leq 30$ μm. Figure 3(a), (b) show the SEM image of MgB$_2$ and p-n junction particles. Figure 3c shows the XRD test curves of pure MgB$_2$, AlGaInP and MgB$_2$+ 1.2wt. % AlGaInP samples, where the vertical dotted line corresponds to the diffraction peak of AlGaInP. The comparison results show that in addition to MgB$_2$ phase, there is an independent AlGaInP phase in the doped sample, indicating that there is no chemical reaction between the two. A certain amount of MgB$_2$ powder raw materials and corresponding inhomogeneous phase p-n particles with different mass fractions were weighed and put into two beakers, respectively, to make alcohol solution and then ultrasonic for 20 min. The two solutions were placed on a magnetic stirrer for stirring, and the inhomogeneous phase solution was added to MgB$_2$ solution drop by drop during stirring. After the dripping, the mixed solution was stirred for 10 min and ultrasonic for 20 min. Then, it was transferred to petri dishes and dried in a vacuum drying oven at 60 ℃ for 4 h to obtain black powder. The powder was fully ground and pressed into a wafer with a diameter of 11 mm and a thickness of 1.2 mm. The pressure and holding time were 14 MPa and 10 min, respectively. The wafer is then placed in a small box made of tantalum, and the box is then placed in an alumina porcelain boat, which is finally transferred to a vacuum tube furnace. In the high pure Ar atmosphere, the samples were slowly heated to 840 ℃ in the vacuum tube furnace for 10 minutes, then cooling to 650 ℃ temperature calcination 1h, and then slowly cooled to room temperature to obtain the corresponding samples [43, 46]. Pure $MgB_2$ samples (represented by S0) and p-n junction doped $MgB_2$ samples were prepared, the concentration of dopant corresponding to each sample is shown in Table 1. In the experiment, the influence of the inhomogeneous phase of luminescent on the superconducting transition temperature of $MgB_2$-based superconductor was studied by changing the content of inhomogeneous phase.

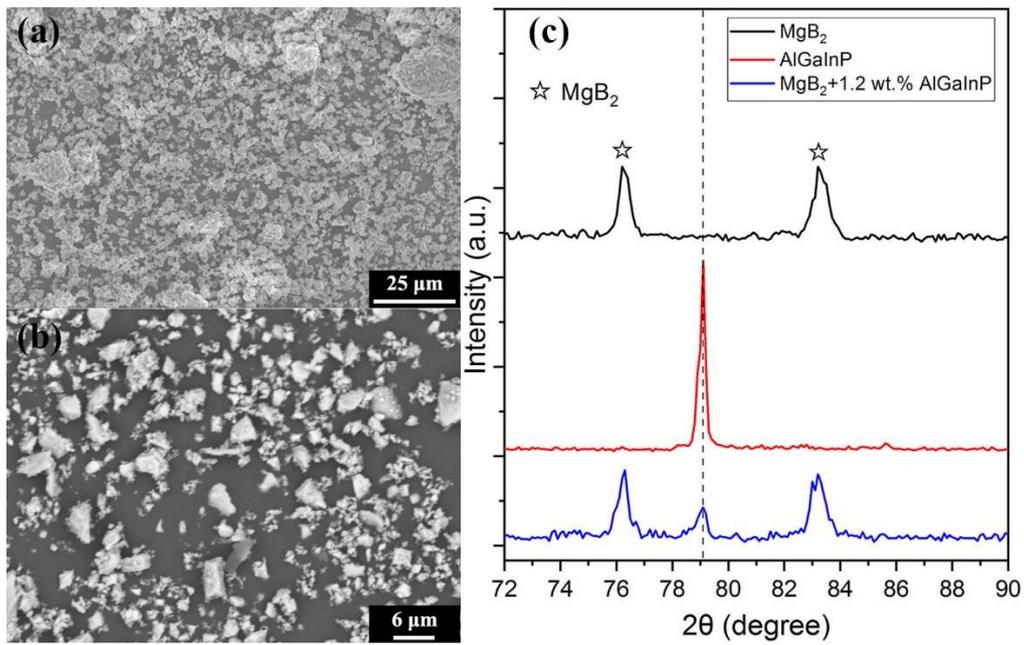

Figure 3. SEM diagram of (a) $MgB_2$ particles, diameter $\Phi \leq 30$ μm, (b) p-n junction particles, ground into 4x4x3μm particles, (c) the XRD test curves of pure $MgB_2$, AlGaInP and $MgB_2$+ 1.2wt. % AlGaInP samples.

Table 1. $MgB_2$ doped inhoumgenous phase with different concentrations

| Sample | S0 | S1 | S2 | S3 | S4 | S5 | S6 |
|---|---|---|---|---|---|---|---|
| inhomogenous phase p-n junction concentration (wt.%) | 0 | 0.5 | 0.8 | 0.9 | 1.0 | 1.2 | 1.5 |

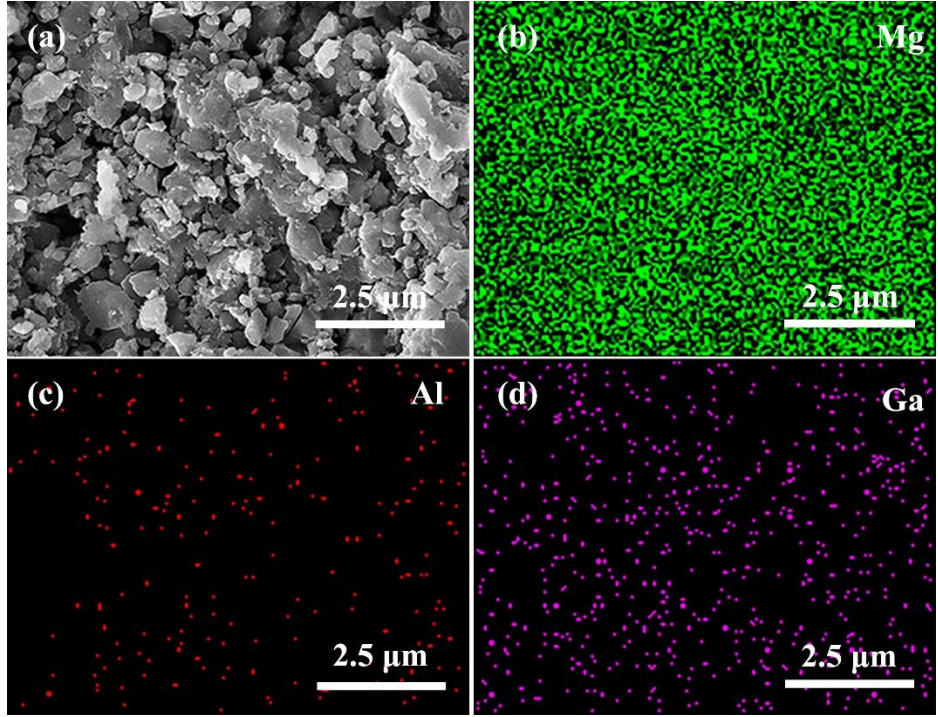

Figure 4. (a) SEM diagram of sample after sintering, (b-d) EDS mapping of Mg, Al, Ga.

Figure 4 (a) shows the SEM image of $MgB_2$ + 0.9% p-n junction after sintering. Figure 4 (b-d) are the EDS mapping for elements Mg, Al, Ga listed in the top right corner of each figure. The distribution of elements in Figure 4 shows the discrete distribution of Al and Ga elements and the aggregation distribution of AlGaInP inhomogeneous phase. There is no chemical intermixing, interdiffusion between $MgB_2$ and AlGaInP.

### *3.3. Critical transition temperature measurement*

A four-lead method was used to measure the R-T curve of the sample at low temperature with a distance of 1 mm between the four probes to determine the superconducting transition temperature Tc of the sample. The closed-cycle cryostat manufactured by Advanced Research Systems provides low temperature environment (minimum temperature is 10 K); the test current (1-100 mA) is provided by the high temperature superconducting material characteristic test device produced by Shanghai Qianfeng Electronic Instrument Co., LTD. Voltages were measured using a Keithley nanovolt meter; adjust the test temperature with a Lake Shore cryogenic temperature controller. The whole testing process is carried out in a vacuum environment [41, 46].

### *3.4. Measurement of critical current density and Meissner effect*

The sample was placed in a low-temperature medium and the current-voltage (I-V) characteristic curve was measured by four-probe method under zero magnetic field. A certain amount of direct current is connected to the prepared sample by two leads, and the other two leads are used to measure the voltage of the prepared sample by a Keithley digital nanovoltmeter. Using indium wire to connect the sample to the lead, and the distance between the two voltage leads for all samples is 1 mm. When the current I passing through the sample exceeds a certain value, the superconducting state is destroyed and changes to the normal state, this current is called the critical transport current of the superconductor. Typically in superconducting systems, the transport critical current density ($J_C$) is determined by I-V measurements at different temperatures (below the initial transition temperature $T_{C,on}$), with a voltage criterion of 1 μV/cm [47, 52-54]. The shape and size of all samples and the distance between the current and voltage leads were kept the same during the test. Subsequently, the prepared samples were tested for DC magnetization [47, 55]. The samples were cooled slowly in a magnetic field of 1.8 mT parallel to the plane, and data were collected during heating. All samples showed complete diamagnetism.

## 4. Results and discussion

Figure 5 is the normalized resistivity curve of the doped x% luminescent inhomogeneous phase p-n junction (x = 0, 0.5, 0.8, 0.9, 1.0, 1.2, 1.5) prepared with $MgB_2$ raw material. x is the doping concentration, where x = 0 is the pure sample $MgB_2$. The black curve in Figure 5a is the normalized R-T curve of pure $MgB_2$ sample, and the results show that the Tc of pure $MgB_2$ sample is 37.4-38.2K. The other six curves correspond to the R-T curve of $MgB_2$ sample doped with p-n junction, and the results show that the Tc corresponding to these six doped samples are 36.8-38 K, 37.4-38.4 K, 37.6-39 K, 37.8-38.8 K, 38-38.7 K and 37.2-38.4 K, respectively. The test results show that at low doping concentrations, such as 0.5%, similar to ordinary chemical doping, the doping of heterogeneous materials reduces the Tc of $MgB_2$ samples (ΔTc < 0) [56,57]. However, when the doping concentration reaches a certain value, such as 0.8%, the inhomogeneous phase enhancement effect occurs, and Tc exceeds the pure sample (ΔTc >0). When the doping concentration is 0.9%, ΔT reaches an increased maximum value of 0.8K and continues to increase the content of inhomogeneous phase, while ΔT decreases instead. The characteristics are the same as those of the previous oxide inhomogeneous phase doping results. [43, 46]

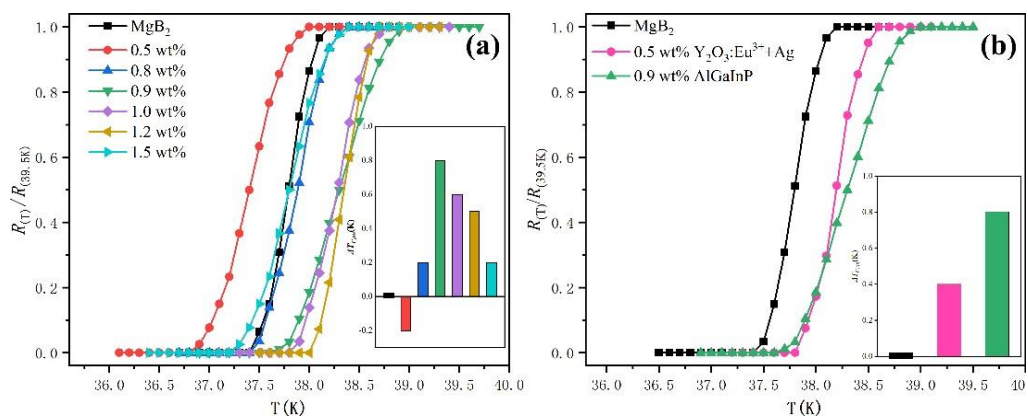

Figure 5. R-T curve of MgB2 sample. (a) Doping results of pure $MgB_2$ and inhomogeneous phase at different concentrations, Insets: the values of $ΔT_c$ ($ΔT_c=T_c−T_{cpure}$). (b) Comparison of the maximum variation of ΔTc produced by pure $MgB_2$ and oxide electroluminescence inhomogeneous

Figure 6 shows the relationship between $J_C$ and temperature of pure $MgB_2$ and p-n junction with different doping concentrations, which is determined by I-V measurement at 20K. It can be seen from Figure 6 (a) that $J_C$ of pure $MgB_2$ and doped samples decreases with the increase of temperature, which is consistent with the results in literature [34, 58, 59]. $J_C$ of pure $MgB_2$ is $8.5 \times 10^4$ A/cm$^2$ at 20K, which is comparable to literature [60, 61]. At this time the $J_C$ of 0.9 wt% luminescent inhomogeneous phase doped sample is $8.83 \times 10^4$ A/cm$^2$. When the temperature is low, $J_C$ decreases slowly, and with the increase of temperature, the decrease speed accelerates. The doping of the electroluminescence inhomogeneous phase increases $J_C$, when T=36 K, $J_C$ of samples with doping concentration of 0.9% is 37% higher than that of pure $MgB_2$. When the inhomogeneous phase concentration is 0.5%, $J_C$ of the sample decreases to a minimum value faster than pure $MgB_2$; and the samples with higher inhomogeneous phase concentration can have $J_C$ at a higher temperature. For example, $J_C$ of pure $MgB_2$ was reduced to a minimum at 38.2K, $J_C$ of 0.9wt% doped sample was reduced to a minimum at 39K, and $J_C$ of 1.2wt % doped sample was reduced to a minimum at 38.7K.

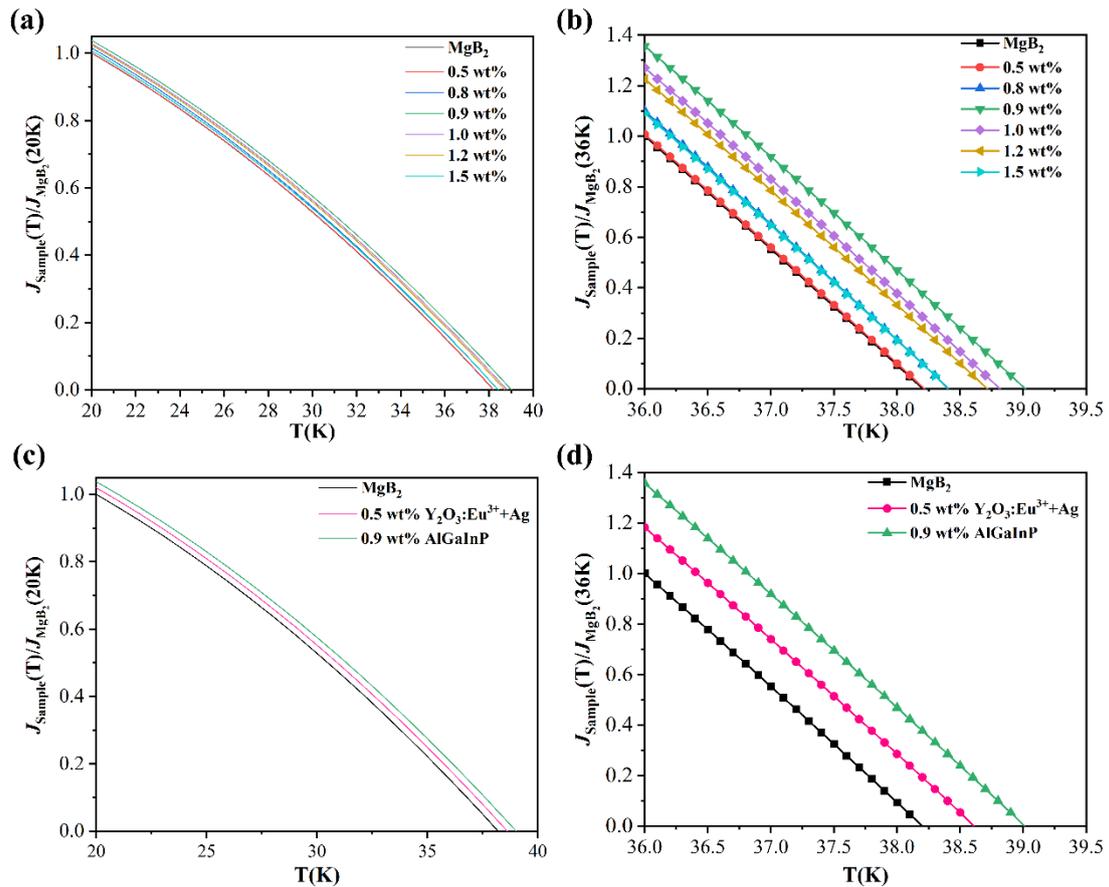

Figure 6 Relationship between $J_C$ and temperature T of $MgB_2$ samples. (a, b) Pure $MgB_2$ and inhomogeneous phase samples. (c, d) Comparison of $J_C$ for the maximum variation of Tc produced by pure $MgB_2$ and oxide luminescent inhomogeneous phase [46], p-n junction inhomogeneous phase.

Figure 7 shows the DC magnetization data of pure $MgB_2$ and $MgB_2$ mixed with inhomogeneous phase. The vertical axis shows the complete diamagnetism change of the material, and the Meissner effect of all samples can be observed through the DC magnetization data. With the increase of temperature, complete diamagnetism, or Meissner effect, weakens and eventually disappears, which is consistent with literature [62-64]. The Meissner effect of pure $MgB_2$ samples disappeared at 37.4K, and that of 0.5% inhomogeneous phase $MgB_2$ samples disappeared at 36.8K. The Meissner effect of 0.9%, 1.0% and 1.2% $MgB_2$ samples disappeared when the temperature was higher than 37.6K and 37.8K and 38K, respectively. It can be seen that the diamagnetic property of the inhomogeneous phase samples with higher concentration is greatly improved compared with that of pure $MgB_2$. For example, the Meissner effect of the 0.9% and 1.2% inhomogeneous phase doped $MgB_2$ samples is increased compared with that of pure $MgB_2$.

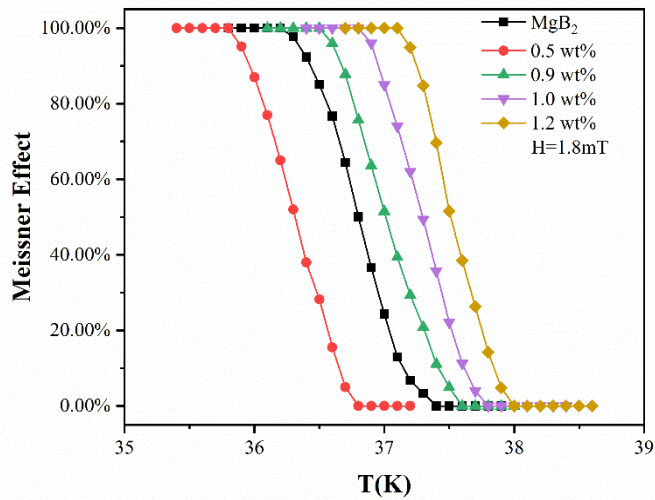

Figure 7. DC magnetization data of pure $MgB_2$ and inhomogeneous phase $MgB_2$.

It can be seen that the p-n junction inhomogeneous phase can produce exactly the same effect as the rare earth oxide inhomogeneous phase, and Figure 5(b) critical transition temperature and Figure 6 (c, d) critical current intensity indicate that the p-n junction inhomogeneous phase's behavior is further improved than that of the rare earth oxide inhomogeneous phase. The main reasons may be:

(1) **The advantages of inhomogeneous phase of p-n junction**     Compared with the previous yttrium oxide inhomogeneous phase, the luminescence intensity and long life of p-n junction are far more than those of the previous electroinduced rare earth luminescence materials (Figure 2), which can ensure the stability of material properties. In addition, p-n junction excitation is easier, only a few volts of excitation, not hundreds or even thousands of volts, the electric field applied by the four-point method for measuring superconductivity can meet the requirement.

(2) **Performance improvement of smart meta-superconductor**     Because p-n junction inhomogeneous phase exhibits good behavior under external field excitation, the optimal amount of dopable inhomogeneous phase increases from 0.5% to 1% compared with oxide. Thus, the critical temperature Tc, current density Jc and Meissner effect of superconducting transition can be improved compared with the oxide inhomogeneous phase, this provides a wider range for the adjustment of material properties. In particular, the performance stability of the material has been greatly improved, and can be stable for more than several hundred hours.

(3)**The origin of smart superconductivity**   In order to explain the above experimental results, we propose an explanatory view of smart superconductivity. Figure 8 shows the schematic diagram of smart superconductivity. By introducing the p-n junction electroluminescent inhomogeneous phase, the photon generated by the inhomogeneous phase in the external field interacts with some superconducting electrons to produce surface plasmon polaritons. The generated evanescent wave can transmit a large number of superconducting electrons with the same energy unimpeded, resulting in the surface plasma system to promote the intense interaction of electrons. The energy injection improves the electron pairing state, promote the superconductivity behavior of the material, increases the critical transition temperature, and forms smart superconductivity.

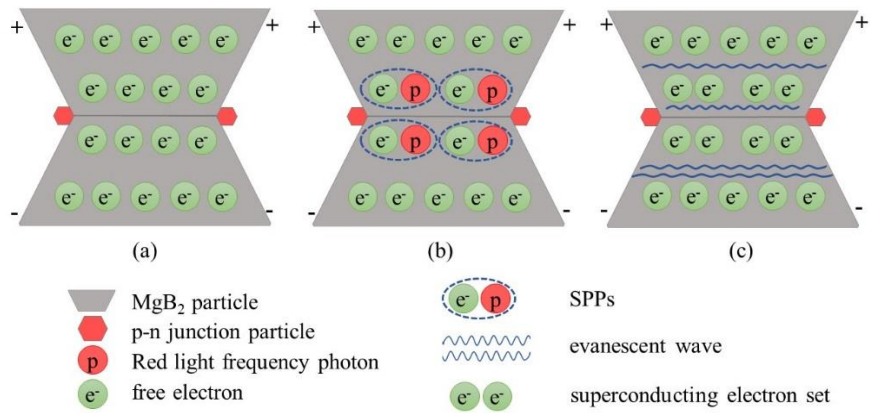

Figure 8. Schematic diagram of smart superconductivity generation. (a) Free electrons are uniformly distributed in the superconductor before the measurement field is applied. (b) After the measurement electric field is applied, the p-n junction particles in red light wavelength glow and generate a large number of photons. The interaction between photons and some conduction electrons occurs at the particle interface, forming plasmons. (c) These surface plasmons propagate as evanescent waves and can transport superconducting electrons. Due to the unimpeded transmission of evanescent wave, the superconductivity of system electrons at higher temperature is promoted.

The conventional superconductor $MgB_2$ is a standard electroacoustic interaction to form a superconducting transition. It can be seen that before reaching the critical temperature Tc, the current density Jc (Figure 6) and the diamagnetism Hc corresponding to Meissner effect (Figure 7) of smart meta-superconductor $MgB_2$ are both higher than that of pure $MgB_2$, it is due to the superposition of the electro-acoustic interaction and the SPPs interaction with superconducting electrons. In the experiment, we used two methods of heating and cooling to test. In the heating method, the temperature of the system is reduced to about 10K first, and then the heating begins.

When the temperature exceeds the critical transition temperature of the pure sample, the electro-acoustic interaction fails, but the interaction between the surface plasmas of the smart meta-superconductor and the superconducting electron still exists. Therefore, the critical transition temperature of the smart meta-superconductor $MgB_2$ is higher than that of the pure $MgB_2$ superconductor (Figure 5). In the state where Jc of pure $MgB_2$ superconductor is zero and diamagnetism disappears, Jc of smart meta-superconductor $MgB_2$ is still non-zero (Figure 6b, d) and diamagnetism still exists (Figure 7). In the cooling method, the system drops from room temperature to 39.1K, and the initial transition occurs as the temperature continues to drop. The temperature has not reached the critical transition temperature of pure samples of 38.2K, and the electro-acoustic effect has not yet taken place, but the superconducting transition phenomenon appears, and Jc of smart meta-superconductor $MgB_2$ also appears. These phenomena confirm that electron-plasmon coupling has been formed, superconducting phase transition has occurred, and smart superconductivity has been achieved. It should be noted that, due to the presence of material resistance, when the temperature reaches a certain point, the interaction between the surface plasmas and the superconducting electrons cannot counteract the effect of material resistance, and superconductivity disappears. However, it is possible to raise the critical transition temperature further by improving the ability of the inhomogeneous phase. Since evanescent waves can exist and transmit at relatively high temperatures, the coupling of superconducting electrons with surface plasmons may promote smart superconductivity at higher temperatures. Here, we find that conventional superconductor $MgB_2$ exhibits smart superconductivity. In fact, in previous studies, we have studied the high-temperature oxide superconductor, and the smart meta-superconductor can also be formed by doping the rare earth oxide inhomogeneous phase [44, 45, 47]. Therefore, it can be concluded that they can also show smart superconductivity by p-n junction inhomogeneous phase, and we will provide the research results in the future.

## 5. Conclusions

In this study, by introducing the p-n junction electroluminescent inhomogeneous phase with red wavelength to realize energy injection, the critical transition temperature Tc, the current density Jc and the diamagnetism of Meissner effect Hc of the smart meta-superconductor MgB are studied. The conclusions are as follows:

(1). Smart meta-superconductor compared with pure MgB2, the critical transition temperature Tc is increased by 0.8K, the current density Jc is increased by 37%, and the diamagnetism of Meissner effect Hc are also significantly improved.

(2). The p-n junction inhomogeneous phase has high luminescence intensity, long stable life and simpler external field requirements. This p-n junction inhomogeneous phase SMSCs produces more significant performance changes than the previous yttrium oxide inhomogeneous phase. Under the same conditions, the critical transition temperature $\Delta T$ increases by nearly 1 times, and the current density Jc increases significantly.

The coupling between superconducting electrons and SPPs is regard as explain this phenomenon. The smart meta-superconductor generated by the inhomogeneous phase opens up a new way to improve the performance of superconductors.